\documentclass[final,5p,twocolumn,number,sort&compress]{elsarticle}
\usepackage{graphicx}
\usepackage{amsmath,amssymb}
\usepackage[utf8]{inputenc}
\usepackage{psfrag}
\usepackage{bm}

\begin{document}
\begin{frontmatter}
\title{Collective strategies and cyclic dominance in asymmetric predator-prey spatial games}

\author[ens,ufrgs]{Annette Cazaubiel}
\ead{annette.cazaubiel@ens.fr}
\author[ufrgs]{Alessandra F. L\"utz}
\ead{sandiflutz@gmail.com}
\author[ufrgs]{Jeferson J. Arenzon\corref{cor1}}
\ead{arenzon@if.ufrgs.br}

\cortext[cor1]{Corresponding author, +55 51 33086478}
\address[ens]{École Normale Supérieure,  International Center of Fundamental Physics, 45 Rue d'Ulm, 75005 Paris, France}
\address[ufrgs]{Instituto de Física, Universidade Federal do
Rio Grande do Sul, CP 15051, 91501-970 Porto Alegre RS, Brazil}

\date{\today}
\begin{abstract}
Predators may attack isolated or grouped prey in a cooperative,
collective way. Whether a gregarious behavior is advantageous to each species depends on
several conditions and game theory is a useful tool to deal with such
a problem. We here extend the Lett-Auger-Gaillard model [Theor. Pop. 
Biol. {\bf 65}, 263 (2004)] to spatially distributed populations and 
compare the resulting behavior with their mean-field predictions for 
the coevolving densities of predator and prey strategies. 
Besides its richer behavior in the presence of spatial organization, 
we also show that the coexistence phase in which collective and
individual strategies for 
each group are present is stable because of an effective, cyclic dominance 
mechanism similar to a well-studied generalization of the Rock-Paper-Scissors 
game with four species, a further example of how 
ubiquitous this coexistence mechanism is. 
\end{abstract}


\end{frontmatter}

\section{Introduction}

There are a myriad of foraging strategies that predators utilize to increase
their success rate. Among them, prey may be attacked in a cooperative, 
coordinated way by a group of predators employing similar spatially and 
temporally correlated actions. When different and 
complementary behaviors are involved, it is also called a collaboration~\cite{BaMyWi13}. Examples of animals that exhibit coordinated or collaborative hunting include lions~\cite{Stander92,HePa95,Legge96} (also the pair of man-eater lions of Tsavo~\cite{Yeakeletal2009}), hawks~\cite{Bednarz88}, crocodiles~\cite{Dinets15}, spiders~\cite{Nentwig85,VaKr01}, ants~\cite{DeLeCoRoCeOrBo10}, and several other species~\cite{BaMyWi13}. 
Interspecies collaborations exist as well, for example,  between fishermen
and dolphins in the south of Brazil~\cite{PrLiLiMi90,DaCaInLuSi12},
honey hunter men and honeyguide birds~\cite{IsRe89,SpBeBe16}, coyotes and badgers~\cite{MiMiLo92},
among others~\cite{BsHoAiFr06}.
Hunting in groups may bring several benefits and has been widely
discussed (for a review, see Ref.~\cite{BaMyWi13} and references therein). For example, it increases the probability of 
capturing large prey~\cite{Bednarz88,CrCr95,Malan98,DeLeCoRoCeOrBo10}, helps prevent the carcass from being
stolen by other predators~\cite{BrBa79,VuPeWa04}, allows for faster food location~\cite{PiMaWi82} and more complex distracting, tracking and chasing tactics, helps related conspecifics that may be unable to hunt or are in the process of acquiring hunting skills~\cite{Malan98,GaGi01}, etc.
On the other hand, there may be setbacks as it also increases the competition between members of the group while feeding, concentrates the search for food to a smaller territory that may decrease the availability of prey, etc. Collective tactics may also benefit prey~\cite{Garay09}. Surveillance is more efficient when done in parallel by several individuals while others have more time to feed themselves~\cite{Pulliam73,Beauchamp08,PaReLoPeGeJa07}. The probability of being caught is smaller~\cite{Hamilton71,IoGoCo12} and the group may take advantage of group  distracting~\cite{CrQu10}, intimidating and escaping techniques. Conversely, a group of prey may be more easily spotted than an individual and the resources
should be shared by all members~\cite{Giraldeau88,Ritz97}. In addition to those factors, for both prey and predators, collective decision making can be improved in larger groups~\cite{Couzin09,CoLi09} (but information sharing may involve costs~\cite{BaWa16} and benefits~\cite{PoSe11}  as well).

Despite mounting experimental results, much less attention has been dedicated to model coordinate hunting~\cite{PaRu88}. Over a decade ago, Lett {\it et al}~\cite{LeAuGa04} introduced a game theoretical model, hereafter referred to as the LAG model, in which the abundance  of prey
and predators were assumed constant and only the fractions of
each populations using either an individual or a collective
strategy coevolved (see, however, Ref.~\cite{McAuLe06}). 
The LAG model takes into account some of the advantages and
disadvantages for both prey and predators choosing a grouping
strategy. More specifically, it is assumed that grouping lowers the
risk of being preyed at the cost of increasing the competition for
resources, while predators have a greater probability of success
at the expense of having to share the prey with others, sometimes
referred to as the ``many-eyes, many mouths'' trade-off~\cite{Giraldeau88, Ritz97}.
Prey and predators were modeled by assuming a fully mixed (no spatial structure), 
mean-field approach, and the temporal evolution of
both densities being described by replicator equations~\cite{HoSi98}.
 
A complementary approach, based on a less coarse grained description,
explicitly considers the spatial distribution of individuals and groups.
The local interactions between them introduce
 correlations that may translate into spatial organization 
favoring either grouping or isolated strategies, raising a number
of questions. For instance, do these strategies coexist within predators or prey populations?  If yes, is this coexistence asymptotically stable?
How does the existence of a local group induce or prevent grouping behavior on neighboring
individuals? Do gregarious individuals segregate,
forming extended regions dominated by groups? In other words, how spatially heterogeneous
is the system?   Does the replicator equation provide a good description for both the dynamics and the asymptotic state? If not, when does it fail?  If many strategies persist,
which is the underlying mechanism that sustains coexistence? 
We try to answer some of these questions with a version
of the LAG model in which space is 
explicitly taken into account through a square lattice whose
sites represent a small sub-population.  Each of the sites
is  large enough to contain only a single group of predators and prey at the same time.
If any of these groups is ever disrupted, their members will resort to a solitary strategy, hunting or defending themselves alone.

The paper is organized as follows. We first review, in Section~\ref{sec.rep}, the LAG model~\cite{LeAuGa04} and summarize the main results obtained with the replicator equation, and then describe, in Section~\ref{sec.space}, the agent based implementation with local competition. The
results obtained in the spatial framework are presented in Sec.~\ref{sec.results}. Finally, we  discuss
our conclusions in Section~\ref{sec.conc}.

\section{The Model}

\subsection{Replicator Equations}
\label{sec.rep}

Lett {\it et al}~\cite{LeAuGa04} considered, within a game theoretical framework, 
grouping strategies for prey and predators. Both can choose between single and
collective behavior
and each choice involves gains and losses for the individuals, as discussed in
the introduction. The relevant parameters of the model are defined in Table~\ref{table.parameters}.
The size of both populations is kept constant during the evolution of
the system; only the proportion of cooperative predators, $x(t)$, and
the fraction of gregarious prey, $y(t)$, evolve in time (see, however,
Ref.~\cite{McAuLe06} for a version that also considers population dynamics). Variations
depend on how the subpopulation's payoff compares with the average payoff of the respective
population. If collective behavior leads to a larger payoff than the average, the associated density increases, otherwise it decreases. This dynamics is
then described by the replicator equations~\cite{HoSi98}.

\begin{table}
{\small
\begin{tabular}{|c|p{7.8cm}|}
\hline
 & \multicolumn{1}{c|}{Definition and default value} \\ \hline
  $p$  &  probability of a predator in a group capturing a lone prey  (0.5) \\
  $G$  &  gain per captured prey per unit of time (1)      \\
  $n$  &  number of predators in a group (3)  \\  
  $e$  &  number of prey captured by a group of predators (2) \\  
  $\alpha$  & preying efficiency reduction due to grouped prey \\ 
  $\beta$ & preying efficiency reduction when hunting alone \\
  $\gamma$ & reduction of prey resources due to aggregation (1) \\
  $F$ & gain for isolated prey per unit of time  (1)\\
\hline           
\end{tabular}
}
\caption{Model parameters~\cite{LeAuGa04} along with 
the default value considered here.}
\label{table.parameters}
\end{table} 
 
For the fraction $x$ of predators hunting collectively, the payoff is~\cite{LeAuGa04}
$$
P_x = \frac{e\alpha pG}{n} y + \frac{pG}{n} (1-y).
$$
The first contribution comes from the interaction of these predators with the
fraction $y$ of prey that organize into groups for defense. By
better defending themselves, prey reduce the hunting efficiency by
a factor $0\leq\alpha< 1$; nonetheless, $e$ prey are captured with probability $p$ and
the gain $G$ per prey is shared among the $n$ members in the group of predators.
The second term is the gain when the group attacks an isolated prey,
whose density is $1-y$, and shares it among the $n$ predators as well.
When the remaining $1-x$ predators hunt solely, they are limited to a single prey and
an efficiency that is further reduced by a factor $0\leq\beta< 1$, what is
somehow compensated by not having to share with others. This information
is summarized in the payoff matrix:
\begin{equation}
A=\left( \begin{array}{cc} 
                       e\alpha p G/n  & pG/n \\
                       \alpha\beta pG & \beta pG 
         \end{array}
  \right).
\label{A}
\end{equation}
As isolated prey consume the available resources, the gain per unit time is, on 
average, $F$. Once aggregated, the resources are shared and
the individual gain reduced by a factor $0\leq\gamma< 1$. The fraction of prey that aggregates 
becomes less prone to be preyed on by a factor $\alpha$. If the grouped prey are
attacked by a group of predators, $e$ prey are captured and 
Lett {\it et al}~\cite{LeAuGa04} considered that the payoff coefficient is
$1-e\alpha p$ (imposing $e\alpha p\leq 1$). On the other hand, a lone predator has its efficiency 
reduced by a factor $\beta$, thus the surviving probability is $1-\beta p$ or
$1-\alpha\beta p$ for an individual or a group of prey, respectively. 
The payoff for the fraction $y$ of prey that remain grouped is then written as 
$$
P_y =  (1-e\alpha p)\gamma F x + (1-\alpha\beta p)\gamma F (1-x).
$$
A similar consideration can be done for isolated prey~\cite{LeAuGa04}, whose payoff matrix is
\begin{equation}
B=\left( \begin{array}{cc} 
                       (1-e\alpha p)\gamma F  & (1-\alpha \beta p)\gamma F \\
                       (1-p)F & (1-\beta p)F 
         \end{array}
  \right).
\label{B}
\end{equation}
It is the difference between the payoff $P$  and its
average, $\overline{P}$, that drives the evolution of
both $x$ and $y$. Indeed, the replicator equations,
$\dot{x}=x(P_x-\overline{P_x})$ and  $\dot{y}=y(P_y-\overline{P_y})$,
which give the rate at which these two densities evolve in time,
are~\cite{LeAuGa04}
\begin{align}
\begin{split}
\frac{\dot{x}}{x}&=(1\ 0) A \left(\begin{array}{c} y\\1-y\end{array}\right)-(x\ 1-x) A \left(\begin{array}{c}y\\1-y\end{array}\right)
 \\
  \frac{\dot{y}}{y}&=(1\ 0) B \left(\begin{array}{c} x\\1-x\end{array}\right)-(y\ 1-y) B \left(\begin{array}{c}x\\1-x\end{array}\right).
  \label{eq.dyn}
\end{split}
\end{align}

These equations describe an asymmetric game and can be rewritten as~\cite{HoSi98}:
\begin{align}
\begin{split}
\dot{x} &= x(1-x)[\alpha_{12}(1-y)-\alpha_{21}y]\\ 
\dot{y} &= y(1-y)[\beta_{12}(1-x)-\beta_{21}x],
\end{split}
\label{dyn}
\end{align}
where
\begin{align}
\begin{split}
\alpha_{12} &= -p(\beta-1/n)G  \\ 
\alpha_{21} &= \alpha p(\beta-e/n)G  \\
\beta_{12}  &= [\gamma-1+\beta p(1-\alpha\gamma)]F  \\
\beta_{21}  &= [1-\gamma-p(1-e\alpha\gamma)]F.
\end{split}
\label{eq.coefficients}
\end{align}
Eqs.~(\ref{dyn}) have five fixed points: the vertices of the unit square, where $x(1-x)=y(1-y)=0$, and the coexistence state:
\begin{align}
\begin{split}
x^* &=\frac{\beta_{12}}{\beta_{12}+\beta_{21}} \\
y^* &=\frac{\alpha_{12}}{\alpha_{12}+\alpha_{21}}.
\end{split}
\label{eq.center}
\end{align}
We will use the notation 01, 11, 10 and 00 either for the asymptotic state  of the whole 
system, i.e., $x_{\infty}y_{\infty}$ with $x_{\infty}\equiv x(t\to\infty)$ and 
$y_{\infty}\equiv y(t\to\infty)$ or, in the spatial version to be discussed
in Section~\ref{sec.space}, for the site variable describing the combined state of site $i$, $x_iy_i$.
The asymptotic state is determined only by the signs of $\alpha_{12}$, $\alpha_{21}$, $\beta_{12}$ and $\beta_{21}$, as 
discussed in Refs.~\cite{HoSi98,LeAuGa04} and summarized here. Indeed, if $\alpha_{12} \alpha_{21}<0$ or $\beta_{12} \beta_{21} <0$, the densities of grouped predators $x$ and grouped prey $y$ 
will monotonically converge to an absorbent state in which at least one of the populations is grouped, i.e, 01, 10 or 11.  On the other hand, if $\alpha_{12} \alpha_{21}>0$ and $\beta_{12} \beta_{21}>0$, there are two different possibilities. From the linear
stability analysis~\cite{HoSi98,LeAuGa04},  $(x^*,y^*)$ is a saddle point if $\alpha_{12} \beta_{12}>0$, and the system ends 
up in one of the vertices of the unit square (for the choice of the parameters considered here, see
below, this does not occur). On the other hand, if $\alpha_{12} \beta_{12}<0$, the eigenvalues are imaginary and the system evolves along closed orbits around the
center $(x^*,y^*)$. In other words, when this last condition is obeyed,  based on a linear stability
analysis, both strategies, grouped or not, coexist at
all times with oscillating fractions of the population and their time averaged
values, once in the stationary regime, correspond to Eqs.~(\ref{eq.center}).

The above replicator equations predict that both $x$ and $y$ approach the asymptotic state 1 or 0 exponentially fast (except in the coexistence phase).  Consider, for instance, phase 11. 
As discussed below, $y$ attains the asymptotic state much faster than $x$. Then, taking $y=1$
and expanding for small $1-x$, we get $x(t)\propto 1-\exp(-\alpha_{21} t)$. The characteristic time is $\tau=\alpha_{21}^{-1}$ and as $\beta\to e/n$, $\tau$ diverges as $\tau\sim (\beta n -e)^{-1}$. Different transitions may depend on other coefficients, 
Eqs.~(\ref{eq.coefficients}), but the exponent of $\tau$ at the transition is always 1.

\begin{figure}[htb]
\includegraphics[width=8.5cm]{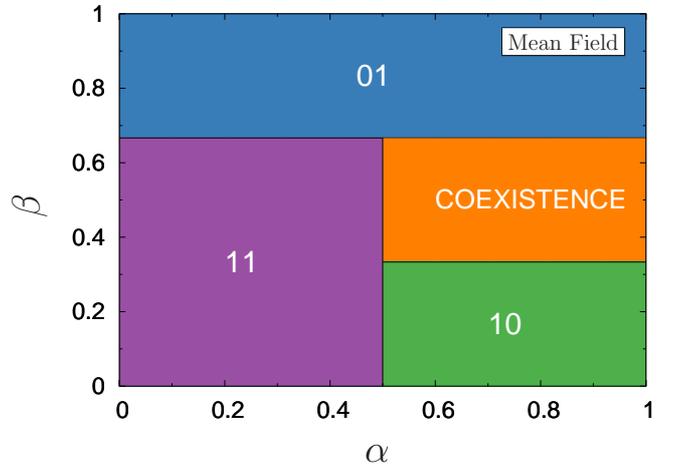}
\caption{Mean-field phase diagram obtained in Ref.~\cite{LeAuGa04}, showing the 11, 01, 10 and coexistence phases. The
vertical transition line is at $\alpha=1/2$ for $0\leq\beta\leq 2/3$ while there is a horizontal
line at $\beta=2/3$, $\forall\alpha$, and another one at $\beta=1/3$ for $\alpha\geq 1/2$. Notice that a coexistence phase replaces phase 00.} 
\label{fig.phasediagram-mf}
\end{figure}

Besides presenting general results, Lett {\it et al}~\cite{LeAuGa04} also discussed the particular case when there is no reduction in resource intake by the prey when they are grouped ($\gamma=1$). In this case, their behavior is a response to the capture rate alone. 
Considering the values listed in Table~\ref{table.parameters} for the several model
parameters ($F=G=\gamma=1$, $p=1/2$, $n=3$ and $e=2$), 
we see that while $\beta_{12}$ is always positive, $\beta_{21}$, $\alpha_{12}$ and $\alpha_{21}$ change signs at $2\alpha=1$, $3\beta=1$ and $3\beta=2$, respectively. These changes in sign lead to different asymptotic behaviors (phases) and locate the transition lines between them,
as can be seen in the mean-field phase diagram of Fig.~\ref{fig.phasediagram-mf}.
 As expected, prey are grouped when $\alpha$ is small, whatever the value of $\beta$. Similarly, small values of $\beta$  lead to cooperating predators for all values of $\alpha$. Remarkably, instead of a 00 phase in which neither species form groups, there is a coexistence phase where the densities of grouped animals oscillate in time along closed orbits around the center point $(x^*,y^*)$ given by Eq.~(\ref{eq.center}). This occurs for $2\alpha>1$ and intermediate values
$1<3\beta<2$, where the eigenvalues of the Jacobian associated with Eqs.~(\ref{dyn}) become purely
imaginary ($\alpha_{12}\beta_{12}<0$, as discussed in 
Refs.~\cite{LeAuGa04,HoSi98}).


\subsection{Spatially Distributed Population}
\label{sec.space}

The above description of the competition between collective and individual strategies
for both predators and prey   does not take into account possible spatial correlations
and geometrical effects. Space is usually introduced by considering an
agent based model in which individuals are placed on a lattice or distributed on a
continuous region. 
We consider the former case with a unit cell corresponding to the size of the territory of the smallest 
viable group. 
Since both predators and prey coexist in each site, there are two variables $(x_i,y_i)$, $i=1,\ldots,N$ 
(where $N=L^2$ is the total number of sites
and $L$ is the linear length of the square lattice) that take only the values 1 or 0; the former when agents are
grouped, the latter for independent individuals. Thus, the global quantities $x(t)$ and $y(t)$ now correspond
to the fraction of sites having $x_i=1$ and $y_i=1$, respectively.

Differently from the mean-field description   where payoff
was gained from interactions with all individuals in the system,  
in the lattice version, interactions are local and occur only between
nearest neighbor sites (self-interaction is also considered since each site
has both predators and prey). At each step of the simulation, one site ($i$)
and one of its neighbors ($j$) are randomly chosen. The predators (prey) on $i$ interact with the prey 
(predators) both on $i$ and in the four nearest neighboring sites, accumulating the 
payoff $P_x^{(i)}$ ($P_y^{(i)}$). At the same time, both groups in $j$ accumulate their payoffs as well. 
If more efficient, the strategies of site $j$ are adopted with a probability proportional to the difference of payoffs. 
For predators (and analogously for prey), this probability is
\begin{equation}
\text{Prob}(x_ i\leftarrow x_j) = \text{max}\left[\frac{P_x^{(j)}-P_x^{(i)}}{P_x^{\text{max}}},0\right],
\label{eq.updating}
\end{equation} 
 where $P_x^{\text{max}}$ is the maximum value of the accumulated payoff of the predators for the chosen   parameters. This rule is known to recover the replicator equation when going from the microscopic, agent based scale to the macroscopic, coarse grained level~\cite{SzFa07}. 
After $N$ such attempts, time is updated by one unit (one Monte Carlo step, MCS). In Ref.~\cite{LuCaAr17}, where some preliminary results were presented, we considered
a slightly different dynamics in which the two chosen neighbors compare their payoff and that earning the smallest one adopted the strategy of the winner (with the above probability). The difference is that, in Ref.~\cite{LuCaAr17}, either $i$ or $j$ would change strategy, while here, only $i$ may be updated.
 
Notice that because the level of spatial description is still larger than the individual
scale, as a first approximation we take
the limit of high population viscosity, neglecting the motion of individuals and groups across 
different sites. In this case, strategies only propagate by the limited dispersal introduced by the
above reproduction (or imitation) rule. We further discuss this point in the conclusions. 
We now present our results for the spatial version of the LAG model and, when possible, compare with
the mean-field predictions.




\section{Results}
\label{sec.results}

\begin{figure}[htb] 
\includegraphics[width=8cm]{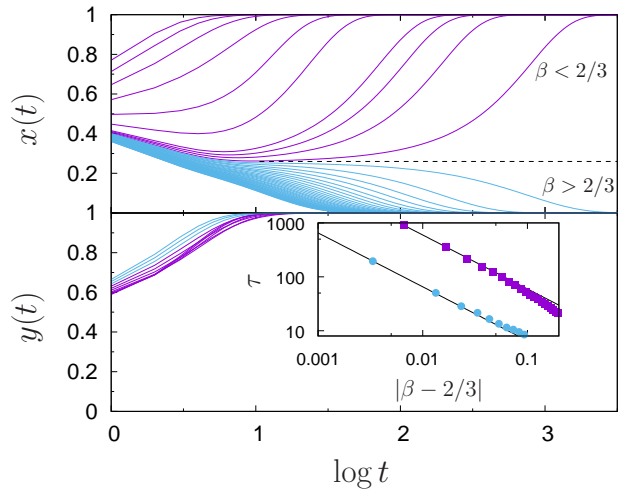} 
\caption{Behavior of both the fraction of predators hunting in groups, $x(t)$, and prey defending themselves collectively, $y(t)$, as a function of time on a single run with $\alpha=0.2$, $L=100$ and several values of $\beta$. The transition from collectively hunting predators to single, individual hunters occurs at $\beta=e/n=2/3$. Notice that most of the prey are in the aggregated state at all times, with $y(t)$ monotonically increasing from $y(0)=0.5$ to $y(\infty)=1$. Predators, on the other hand, have a richer behavior (see text). As $\beta\to 2/3$, from both sides, a plateau at $x\simeq 0.26$ develops (the dashed, horizontal line is only a guide to the eyes). Inset: Power-law behavior of the characteristic time $\tau$ such that $|x-x_{\infty}|<0.2$ around the transition at $\beta=2/3$. The straight lines have exponent 1, as predicted by the replicator equation, although the coefficients differ by one order of magnitude. The top curve is for $\beta\to 2/3^-$ while the bottom one is for $\beta\to 2/3^+$.
}
\label{fig.xy_a02}
\end{figure}

Fig.~\ref{fig.xy_a02} shows the temporal evolution of both $x(t)$ and $y(t)$ for $\alpha=0.2$,
several values of $0<\beta<1$ and an initial state with $x(0)=y(0)=1/2$.
In this case, for all values of $\beta$, prey remain mostly grouped at all times since $y(t)>0.5$ (bottom panel). As predicted in mean-field,
there is a transition at $\beta=2/3$ where predators change strategy (top panel): for $\beta>2/3$, hunting alone becomes more efficient and
 $x_{\infty}=0$. On the other hand, when the cost of sharing the prey is compensated by more efficient preying, $\beta<2/3$, we have $x_{\infty}=1$.  Interestingly, for the initial state chosen here, the behavior of $x(t)$ is not monotonic when 
$(1+2\alpha)/3(1+\alpha)<\beta<2/3$:
$x(t)$ initially decreases, $\dot{x}(0)<0$, attaining a minimum
value and then resumes the increase towards $x_{\infty}=1$. The location of this minimum 
corresponds to the time at which $y$ crosses the point $y^*$, and the envelope of all
minima follows the plateau developed for $\beta>2/3$ as this value is
approached from above. In this latter region, the behavior follows a two-step
curve: there is a first, fast approach to the plateau followed by the departure from it at a much
longer timescale. 
In this model, the fast relaxations occur as prey organize themselves into groups  (increasing $y$) while the later slow relaxation is a property of the
predators alone and is caused by the orbit passing near an unstable fixed point, as 
can be understood from Eqs.~(\ref{dyn}). For values of $\alpha$ in the interval $(1+2\alpha)/3(1+\alpha)<\beta<2/3$, if $y<y^*$, $x$ decreases until $y$ crosses the line at $y^*$, whose value depends on both $\alpha$ and $\beta$. At this point, the coefficient of $\dot{x}$ is zero and there is a minimum. 
Once $y>y^*$, $x$ resumes the increase and approaches the asymptotic state $x_{\infty}=1$
exponentially fast. As $\beta\to 2/3^-$, $y^*\to 1$ and the minimum
crosses over to an inflection point. Indeed, for $\beta>2/3$, $x(t)$ decreases
towards 0 after it crosses the plateau. This behavior is seen both in simulations and in mean-field.

Associated with the late exponential regime there is a characteristic time that diverges as a phase transition is approached. For example, in mean-field, $\tau\sim \alpha_{21}^{-1}\sim (\beta n-e)^{-1}$ for the 11-01 transition. In the simulations, as $x$ approaches the limiting value $x_{\infty}$, 
$\tau$ is estimated as the time beyond which $|x-x_{\infty}|<\epsilon$, where $0<\epsilon<1$ is chosen, for convenience, to be $\epsilon=0.2$.  The exponent measured in the simulations is in
agreement with the mean-field prediction, as can be seen in the inset of Fig.~\ref{fig.xy_a02}.

\begin{figure}[htb]  
\includegraphics[width=8cm]{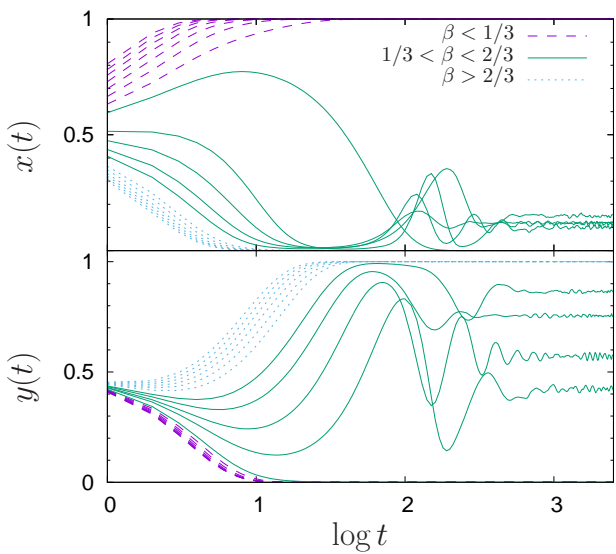} 
\caption{The same as in Fig.~\ref{fig.xy_a02} but for $\alpha=0.8.$ Two transitions are present at $\beta=1/n=1/3$ and $e/n=2/3$, with a coexistence phase in between where both $x_{\infty}$ and
$y_{\infty}$ attain a plateau.}
\label{fig.xy_a08}
\end{figure}

\begin{figure}[htb]
\includegraphics[width=8.5cm]{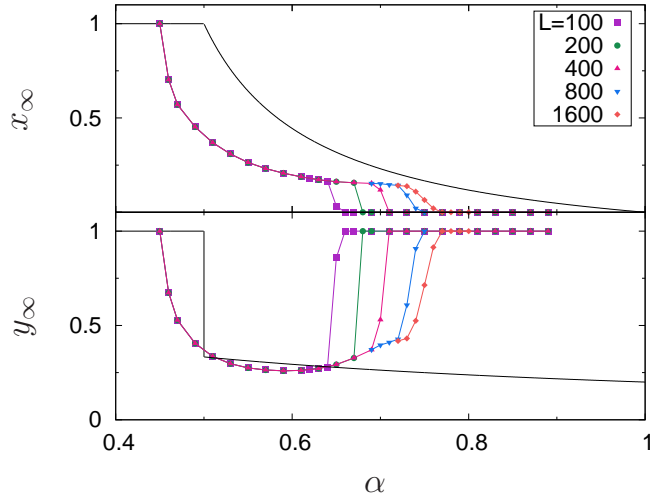} 
\caption{Average asymptotic value of the fraction of predators hunting in group, $x(t)$ (top), and prey defending
themselves in group, $y(t)$ (bottom),
for $\beta=0.4$, as a function of $\alpha$. The solid lines correspond to the fixed points, Eqs.~(\ref{eq.center}), predicted in mean-field and discussed in Sec.~\ref{sec.rep}.  Notice that the transition from the 11 phase to the coexistence one is continuous for both quantities and occurs at a smaller value of $\alpha$ than predicted in mean-field, $\alpha_c\simeq 0.45$. At a larger, size dependent value of $\alpha$ there is a transition to the 01 phase. As the system size $L$ increases, the coexistence phase becomes broader. Averages are over 10 samples.}
\label{fig.asympt.xy}
\end{figure}

Differently from the $\alpha=0.2$ case, for $\alpha=0.8$ there is, in agreement 
with the replicator equation, a coexistence phase where both strategies may persistently coexist. 
Fig.~\ref{fig.xy_a08} shows that, depending on $\alpha$ and $\beta$, there is an initial, transient regime
in which both $x$ and $y$ oscillate, 
getting very close to 
the absorbing states.
Because of the stochastic nature of the finite system, it may eventually end in one of those absorbing states during the oscillating regime (although the required time may be exponentially large, see later discussion),
otherwise the amplitude of the oscillations decreases and a mixed fixed point is attained. A natural question is how close this fixed point is to the mean-field prediction, Eqs.~(\ref{fig.asympt.xy}). 
Since it is the intermediate region in which coexistence may occur that presents new, non trivial behavior, we now discuss it in more detail. 


\begin{figure}[htb]
\includegraphics[width=8.5cm]{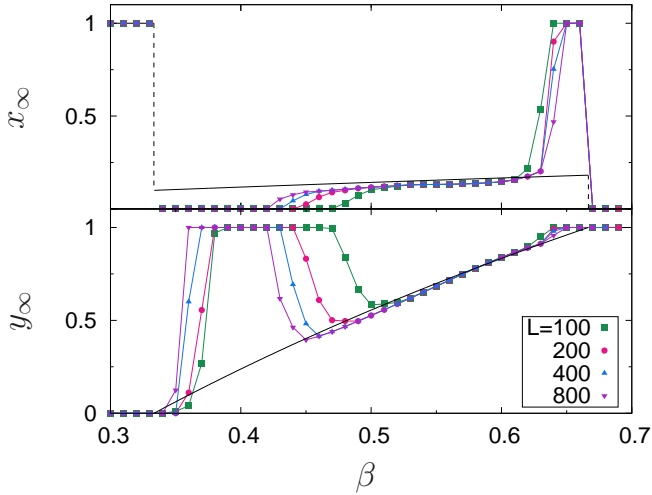} 
\caption{The same as Fig.~\ref{fig.asympt.xy} but with $\alpha=0.8$ and $\beta$ as the variable. Averages are over 10-20 samples, depending
on the size.}
\label{fig.asympt.xya8}
\end{figure}

\begin{figure}[htb]
\includegraphics[width=8.5cm]{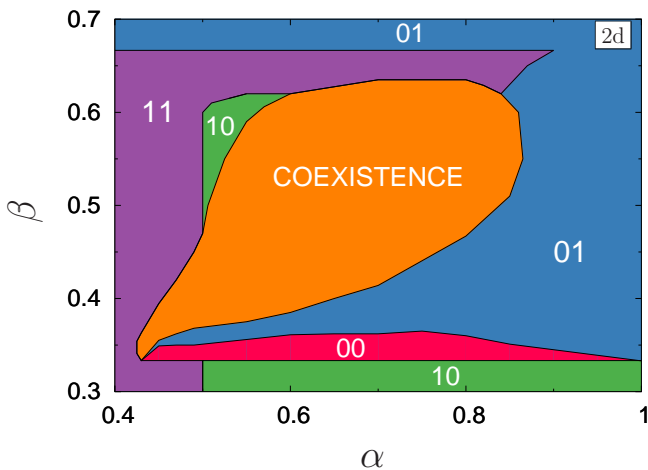}
\caption{Zoomed in phase diagram, obtained with $L=100$, showing the rich behavior when agents are spatially distributed (compare with the pure
coexistence phase existing in the region $1/2\leq\alpha\leq 1$ and
$1/3\leq\beta\leq 2/3$ in the mean-field diagram, Fig.~\ref{fig.phasediagram-mf}). 
As $L$ increases, the coexistence region gets bigger while the phases around it shrink (in
particular the small ones like 10 and
00). Whether they disappear or remain very small when $L\to\infty$ is not clear.
} 
\label{fig.phasediagram}
\end{figure}

The mean-field monotonic behavior can be observed in the solid lines of Fig.~\ref{fig.asympt.xy} as the system enters the coexistence region at a fixed $\beta=0.4$ from the 11 phase. While mean-field predicts that $x_{\infty}$ and $y_{\infty}$ present a continuous and discontinuous transition, respectively, at $\alpha_c^{\scriptscriptstyle\rm MF}=0.5$, when predators and prey are spatially distributed, the simulation shows, instead, continuous transitions for both quantities at a smaller value, $\alpha_c\simeq 0.45$. This is not due  to finite size effects since the curves, for the broad range of system sizes considered here, collapse onto a single curve in this region, and the larger the system is, the wider the collapsed region becomes. 
The simulation results on the lattice disagree with the mean-field predictions both quantitative and
qualitatively.
In the spatial model, instead of a monotonic decay, $y_{\infty}$ has a minimum at $\alpha\simeq 0.59$ for $\beta=0.4$, even in the limit of very large systems.  The existence of this minimum  is remarkable because one would expect that as $\alpha$ increases, predators become more efficient against grouped prey and the fraction of the latter would decrease. 
Moreover, both $x_{\infty}$ and $y_{\infty}$ present strong finite size effects in this phase. Above a size dependent value of $\alpha$, $x$ is absorbed onto the group disrupted state ($x_{\infty}=0$) and, immediately after, $y_{\infty}$ evolves toward 1. Fig.~\ref{fig.asympt.xya8} also shows the behavior of $x_{\infty}$ and
$y_{\infty}$, but for a fixed value of $\alpha$. In the interval $1/3<\beta<2/3$, coming from the
10 phase, the system first goes through the 00 and 01 phases (the former is not present in the mean-field case),  before crossing the wider coexistence region. Then it crosses region 11 and eventually arrives again
at the 01 phase when $\beta=2/3$. All these behaviors and transitions  are summarized in the phase 
diagram of Fig.~\ref{fig.phasediagram}, for $L=100$. The coexistence region, that is reentrant and exists also for $\alpha<1/2$, increases with the system
size (as can be observed in Figs.~\ref{fig.asympt.xy} and \ref{fig.asympt.xya8}). The small regions  shrink for increasing $L$, but the convergence to the $L\to\infty$ behavior (e.g., whether 00 and 10 phases completely disappear or not), is very slow.
A more detailed analysis can be done (see Refs.~\cite{AnSc06,ReMoFr07,LuRiAr13} and references therein) by measuring the time it takes for the coexistence state to be absorbed into
one of the homogeneous states (what eventually will occur for a finite system because of the stochastic nature of the dynamics) and how it depends on $L$. Indeed, inside the coexistence region, this characteristic time
increases exponentially  with the system size while it is much smaller (logarithmically) for the
other phases.

\begin{figure}[htb]
\includegraphics[width=2.7cm]{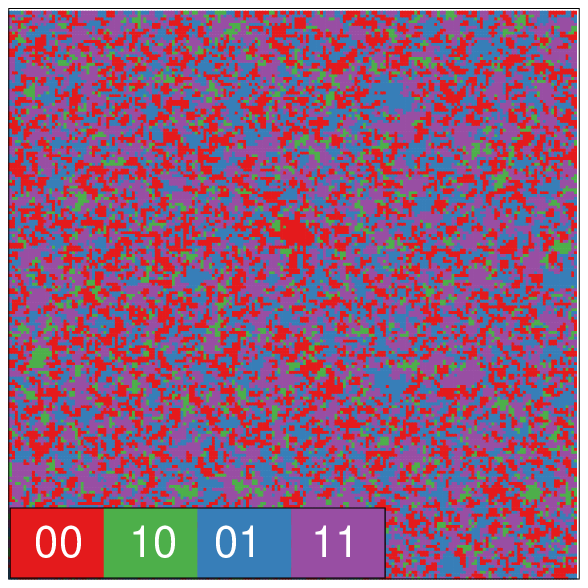}
\includegraphics[width=2.7cm]{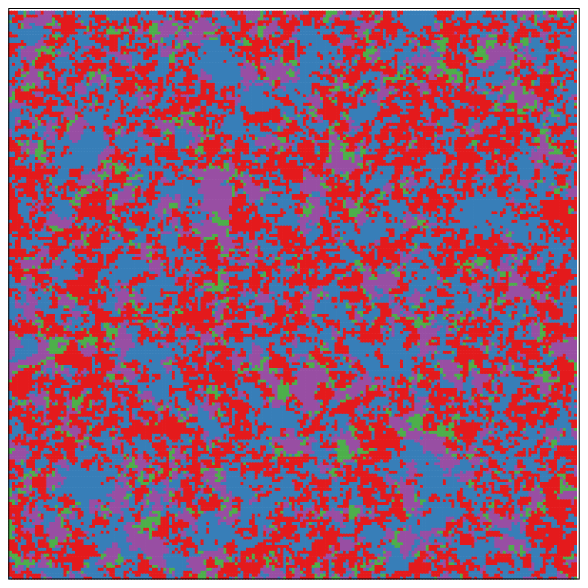}
\includegraphics[width=2.7cm]{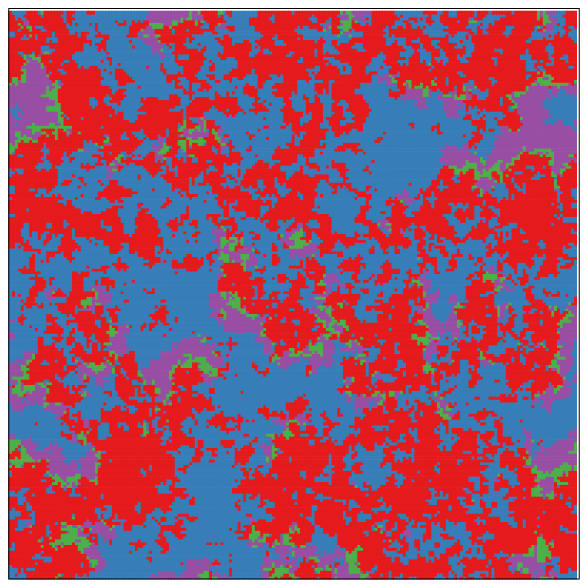}
\begin{center}
(a) \hspace{2cm} (b) \hspace{2cm} (c)
\end{center}
\caption{Snapshots using a color code for the combined $x_iy_i$ strategy for $\beta=0.5$ and $\alpha=0.6$ (left), 0.7 (middle) and 0.8 (right), all inside the coexistence phase, at $t=2^{13}$ MCS and on an $L=200$ lattice. Notice that the strategies organize in intertwined domains, whose characteristic size depends on $\alpha$ and $\beta$. In this case, both 00 and 01 increases as the system approaches the border with the 01 phase.} 
\label{fig.4snapshots}
\end{figure}

\begin{figure*}
\includegraphics[width=10.5cm]{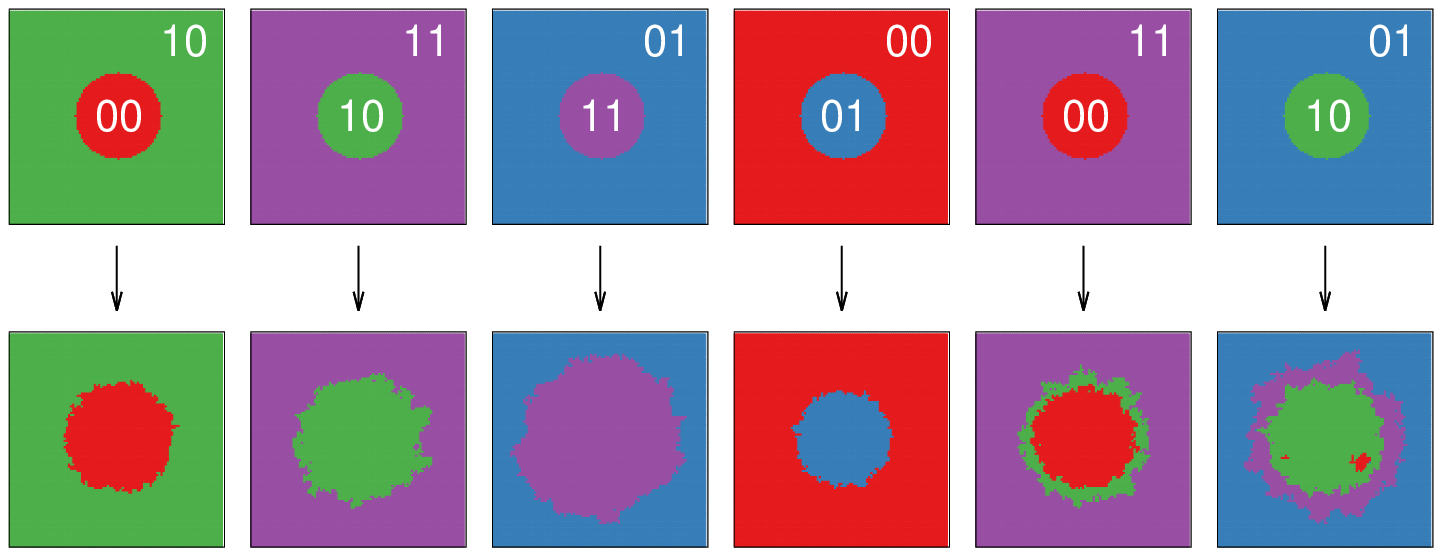}
\hspace{1cm}
\includegraphics[width=4cm]{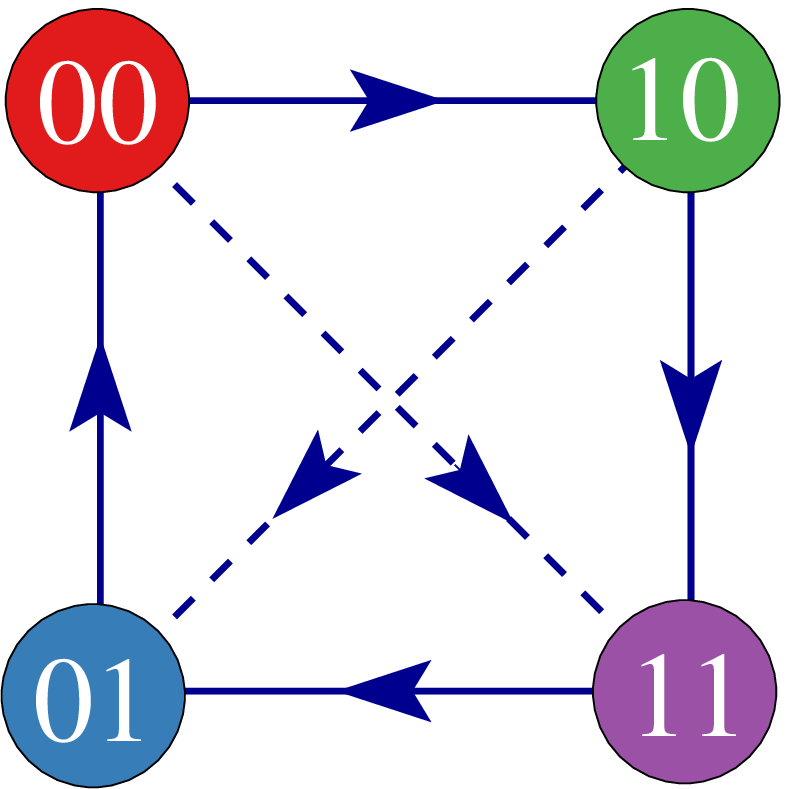}
\begin{flushright}
(a)\hspace{8cm}(b)\hspace{4cm}\mbox{}
\end{flushright}
\caption{a) Strategy dominance for $\beta=0.4$ and $\alpha=0.7$ starting from an initial, circular patch (top row) and evolving for 40 MCS (bottom row). In some cases, denoted by the solid arrows in the graph at the right, the initial patch increases in size. For the interactions along the diagonal (dashed arrows), the invasion occurs in two steps. For example (fifth column), 00 first disrupts the aggregated prey, forming an intermediate 10 cluster, and then grows into it. Notice that for the 10 invading 01 (last column), besides the strategy 11 that intermediates the invasion, there are some groups of 00 growing inside the 10 patch (at the interface 01-10, all combinations may be created and, sometimes, migrate toward the interior of the circle).  b) Interaction graph showing the direction of the invasion front for each of the four strategies.}
\label{fig.circ_invasion}
\end{figure*}

In the coexistence phase, any of the four combinations of lone and collective strategies for both predators and prey may be present at each site ($x_iy_i$): 00, 01, 10 and 11. Fig.~\ref{fig.4snapshots} shows, for $\beta=0.5$ and different values of $\alpha$, how these four strategies organize into domains whose sizes depend on both $\alpha$ and $\beta$. 
While the coexistence persists, the four strategies survive by spatially organizing themselves in nested domains whose borders move, invading other domains in a cyclic way. 
In Fig.~\ref{fig.circ_invasion}a we show the direction of these invasions by placing, in the initial state ($t=0$, top row), one strategy inside, and another outside a circular patch (a flat interface would do as well). The bottom row shows the corresponding state after 40 Monte Carlo steps. By swapping positions, the invasion direction is reversed, indicating that it is not a simple curvature driven dynamics but, instead, involves a domination relation. Taking into account the six combinations of
Fig.~\ref{fig.circ_invasion}a, the interaction   graph shown in Fig.~\ref{fig.circ_invasion}b summarizes how the different strategies interact (this particular orientation of the arrows may change for other points in the coexistence phase~\cite{LuCaAr17}).
Similar cyclic dominance behavior has been extensively studied in predator-prey models with interaction graphs with three  (Rock-Paper-Scissors) or more species~\cite{SzFa07,SzMoJiSzRuPe14} with intransitive (sub-)loops. Indeed, the topology of the interaction   graph in Fig.~\ref{fig.circ_invasion}b is closely related to the one in Ref.~\cite{LuRiAr13}. The invasion is either direct as in the first four columns (or, equivalently, along the perimeter of the interaction graph) or, as seen in the last two columns, involves the creation of an intermediate domain. For example, a patch of 00s invades 11 (fifth column in Fig.~\ref{fig.circ_invasion}a) by first disrupting the prey organization. The strategy 10 thus created, invades 11 and, in turn, is invaded by 00. In those cases that the invasion proceeds through an intermediate domain (strategies along the diagonal of Fig.~\ref{fig.circ_invasion}b are not neutral), we use a dashed arrow in the interaction graph. This cyclic dominance among the combined strategies of prey and predators is the  mechanism underlying the persistence of the coexistence state in this region of the phase diagram.

While the solid lines shown in the interaction graph of Fig.~\ref{fig.circ_invasion}b are valid 
under broader conditions, the diagonal, dashed ones are not.  For example,
if only 00 and 10 sites are present (Fig.~\ref{fig.circ_invasion}a, first column), prey will never change to a collective strategy (11 and
01 sites will not appear). Only predators may change strategy in this case and because they earn their payoffs
from isolated prey, this is easy to evaluate and does not depend on the specific geometry of either the
lattice or the initial configuration. Indeed, for a 00 site comparing its payoff with a
10 neighbor, while the former earns $5\beta/2$, the latter receives 5/6. The 00 strategy will then 
invade the 10 population if $\beta>1/3$. If we repeat this analysis for the first four 
columns of Fig.~\ref{fig.circ_invasion}a,
we get: 10 invades 11 if $\alpha>1/2$, 11 invades 01 if $\beta<2/3$ and, finally,
01 invades 00 if $\alpha<1$. These are the interactions around the graph 
of Fig.~\ref{fig.circ_invasion}b, represented by solid lines. Thus, as expected, by changing the values of $\alpha$ and $\beta$, the 
arrows may change orientation as well. Indeed, the possible combinations of these arrows agree
with the regions shown in the mean-field phase diagram of Fig.~\ref{fig.phasediagram-mf}. In
particular, the one with $\alpha>1/2$ and $1/3<\beta<2/3$ is where the four arrows are clockwise oriented,
dominance is cyclic and coexistence may ensue. Diagonal interactions are much more complex. All four states may now be
created from initial states with only 00 and 11 (or 10 and 01) and the earned payoffs
depend on the number of neighbors having each strategy, what changes from site to site and evolves in 
time, leading to the rich phase diagram inside the coexistence region, shown in Fig.~\ref{fig.phasediagram}. Indeed, it is important to emphasize that the invasion graph in Fig.~\ref{fig.circ_invasion}b was obtained for the particular values $\beta=0.4$ and $\alpha=0.7$ within that time window.  Other points inside the coexistence phase may show similar behavior albeit with some arrows reversed and strategies switching roles while keeping some of the sub-loops intransitive and the 
coexistence stable~\cite{LuCaAr17}.
An intriguing characteristic of the phase diagram, Fig.~\ref{fig.phasediagram}, is that the
coexistence phase is surrounded by several smaller phases, absent in that region within mean-field.
This behavior can be better understood observing the patterns in Fig.~\ref{fig.4snapshots}. 
In that case, as $\alpha$ increases and $\beta$ is kept fixed, i.e., one approaches the frontier with the 01
phase, the densities of both 00 and 01 strategies increase (remember that 01, as shown in 
Fig.~\ref{fig.circ_invasion}b, prey on 00). There is, for a given size $L$, a value of $\alpha$
where those domains are comparable to $L$ and it seems that this finite size effect triggers
the transition, originating the strong dependence of its location on $L$. 
If we move toward the other phases,
a similar behavior occurs, with the dominant strategy of the approaching phase increasing
in density. The important point is that, as the system increases, the coexistence phase becomes larger and the other phases
decrease in size. Whether this trend eventually leads to a coexistence phase occupying the whole region is an open question, beyond our current computational capabilities.

\section{Conclusions and discussion}
\label{sec.conc}

The foraging behavior of predators and the corresponding defensive response from prey is fundamental to understand how small, spatially distributed animal communities organize and eventually engage in more complex forms of sociability~\cite{VeGaAv10,PoLe14,Read16}. It thus becomes important, when modeling such behavior, to go beyond mean-field where a fully mixed, infinite size system is considered and the spatial structure, with the correlations it implies, is missing. We considered here a finite dimensional stochastic version of the model introduced by Lett {\it et al}~\cite{LeAuGa04} in which short range interactions between predators and prey are taken into account. In this way, local spatial correlations that change, to some extent, the foraging behavior predicted by the replicator (mean-field) equation, and the accompanying new dynamical behavior are introduced. The game theoretical framework in this model considers two strategies, collective or individual, for both hunting predators and defensive prey. The advantages and disadvantages of each option are modeled by a set of independent parameters from which we considered variations in only two: how the probability $p$ of a group of predators capturing
a lone prey is reduced when prey are grouped ($\alpha p$) or when the predators hunt alone ($\beta p$). We then study in detail this particular case, discussed only within mean-field in Ref.~\cite{LeAuGa04}, in the limit of high population viscosity in which strategy dispersion is a slow process, solely occurring due to the newborn limited dispersal driven by the pairwise updating rule between nearest neighbors patches. 

The combined strategies of both predators and prey present four possible states in our binary version. There are three phases with either prey or predators (or both) behaving collectively (10, 01 and 11), while in the 00 phase there are no groups. Results from extensive simulations on the square lattice
confirm part of the mean-field phase diagram, with a richer structure around the coexistence phase. More importantly, our results unveil the underlying mechanism for the coexistence of these strategies. Specifically, in this phase, this model is an example of an asymmetric game presenting cyclic dominance between the above four combined strategies. Starting, say, with a population of collective predators preying on lone individuals (i.e., most of the sites are 10 while the other strategies, albeit present, have small densities), free riders have the advantage of not having to share their prey with the other members of the group and the 00 strategy invades 10. At this point, if prey organizes into groups they may better defend themselves, and thus 01 replaces 00. Finally, the lone predators get better off preying collectively and 11 dominates 01. Finally, 10 invades 11 because 
it is better for predators to go after grouped prey, since the number
of captured prey ($e$) compensates for the loss of efficiency
($\alpha$). Thus, it is better
for aggregated prey to stay alone.
The strategies 11, 01, 11 and 00 form spatially extended domains whose borders are not static and move accordingly with the interaction graph (e.g., Fig.~\ref{fig.circ_invasion}b). 

Besides considering simple models for this less explored foraging variability, our results emphasize the importance of studying both the asymptotic states and the dynamics towards them in a finite size population.
It is not possible to disentangle the strong finite size effects observed in
the asymptotic state of the stochastic model from the dynamical evolution since it is the very existence of 
orbits that, by closely approaching the absorbing states, makes the system prone to be captured by them as a consequence of fluctuations. Since actual populations are  far from the thermodynamical limit (infinite population), the results obtained for
intermediate sizes become relevant. Timescales may be very large and the transient coexistence may extend to times much larger than those that are relevant in practice. Moreover, the differences between the mean-field and the spatial version show the importance of trying several complementary approaches even when studying quite simplified models.  In our simulations, the microscopic updating rule, 
Eq.~(\ref{eq.updating}), is applied to one site $i$ chosen at random, i.e., both $x_i$ and $y_i$ may be updated in each single step. We checked that the results are essentially the same when they are
independently chosen or, as in Ref.~\cite{LuCaAr17}, when either $i$ or $j$ change at any given attempt. In this regard, it would be important to further check how robust the results are when also changing
the updating rule, replacing Eq.~(\ref{eq.updating}) by, for example, the Fermi rule (see Ref.~\cite{SzFa07} for a review on the
several possible dynamics).

Prey and predators usually engage in somewhat coordinated chase and
escape interactions~\cite{KrRe96,Nahin07,CoPlBe08,GiBa10,RaBaLuRiSoVi11,Angelani12} that also allow them to explore and profit from neighboring patches. 
It is thus interesting to check whether, and to what extent, the properties of the model, in particular in the coexistence region, change in the presence of mobile individuals~\cite{ReMoFr07}. 
Chasing and escaping behaviors may be quite complex depending on the physical and cognitive capabilities of each individual and involve space and time correlations between their displacements and changes in velocity~\cite{ViZa12}. Simple movement rules, if not completely random, are likely to generate repeatable (and, because of that, exploitable) patterns of behavior, while those involving higher levels of variability and complexity somehow involve more advanced cognitive skills. By studying such mobility patterns, one could get a better understanding of how important those cognitive abilities are in defining hunting strategies~\cite{BaMyWi13}. Moreover, these patterns are also relevant for the demographic distribution of both predators and prey since the shuffling of strategies, depending on how random the mobility is~\cite{VaSiAr07}, may decrease the spatial correlation and destroy local structures, changing the spatial organization of prey and predators~\cite{PoLe14}.  

Several other extensions of this model are possible. For simplicity, we assumed that the size of a group is constant and homogeneous throughout the population. However, it can also be  considered as a dynamical parameter coevolving together with the collectivist trait. It is possible to further explore the possibilities offered by the spatial setup, for example, having heterogeneous parameters depending on the landscape
(due to an uneven distribution of resources or to the variability of the species), what may induce an optimal, intermediate level of collective foraging corresponding to mixed strategies~\cite{ScKi13,BhVi14,BhVi15}, sometimes hunting in group, sometimes alone.
Another important question is
how the size of the hunting groups respond to an increase or to stochastic fluctuations
in the size of a swarm of prey (and vice-versa). Although we focus here on the discrete, binary 
situation, it is also possible to have larger patches with enough individuals to form
more than one group, such that the variables describing the local populations may have multiple allowed
values or even become continuous. In addition, the dynamics considered here allows neither a 
species with a smaller payoff to be the winner nor the reintroduction of an already extinct species.
Removing such constraints (what can be considered a kind of external noise) may be an effective way of
taking into account some of the missing ingredients of the model.
It would thus be interesting to see how robust the results are in the presence of such noise. Finally, the short-range interactions present in the model are not able to synchronize well separated regions  in the coexistence phase. On the contrary, mean-field~\cite{SaTo94,LeAuGa04} predicts a neutral fixed point in the coexistence region, around which the system oscillates.   By introducing a fraction of long-range connections we expect such global oscillations to be restored~\cite{KuAb01,RuAr14}. What is the threshold fraction of such interactions for having global oscillations and how it depends on the parameters of the model, along with the other points raised above, are still open questions.

\chapter*{Acknowledgments}
AC and AFL equally contributed to this work. We thank Mendeli Vainstein for a critical
reading of the manuscript.
AC thanks the IF-UFRGS for the hospitality and the ENS-Paris for 
partial support during her stay in Porto Alegre. AFL is partially supported by a CNPq PhD grant.
JJA thanks the INCT Sistemas Complexos and the Brazilian agencies CNPq, Fapergs and 
CAPES for partial support.



\begin{thebibliography}{10}
\expandafter\ifx\csname url\endcsname\relax
  \def\url#1{\texttt{#1}}\fi
\expandafter\ifx\csname urlprefix\endcsname\relax\def\urlprefix{URL }\fi
\expandafter\ifx\csname href\endcsname\relax
  \def\href#1#2{#2} \def\path#1{#1}\fi

\bibitem{BaMyWi13}
I.~Bailey, J.~P. Myatt, A.~M. Wilson, Group hunting within the carnivora:
  physiological, cognitive and environmental influences on strategy and
  cooperation, Behav. Ecol. Sociobiol. 67 (2013) 1--17.

\bibitem{Stander92}
P.~E. Stander, Cooperative hunting in lions: the role of the individual, Behav.
  Ecol. Sociobiol. 29 (1992) 445--454.

\bibitem{HePa95}
R.~G. Heinsohn, C.~Packer, Complex cooperative strategies in group-territorial
  {A}frica lions, Science 269 (1995) 1260--1262.

\bibitem{Legge96}
S.~Legge, Cooperative lions escape the prisoner's dilemma, Tree 11 (1996) 2--3.

\bibitem{Yeakeletal2009}
J.~D. Yeakel, B.~D. Patterson, K.~Fox-Dobbs, M.~M. Okumura, T.~E. Cerling,
  J.~W. Moore, P.~L. Koch, N.~J. Dominy, Cooperation and individuality among
  man-eating lions, Proc. Nat. Acad. Sci. 106 (2009) 19040--19043.

\bibitem{Bednarz88}
J.~C. Bednarz, Cooperative hunting in {H}arris' hawks, Science 239 (1988)
  1525--1527.

\bibitem{Dinets15}
V.~Dinets, Apparent coordination and collaboration in cooperatively hunting
  crocodilians, Eth. Ecol. Evol. 27 (2015) 244.

\bibitem{Nentwig85}
W.~Nentwig, Social spiders catch larger prey: a study of \textit{{A}nelosimus
  eximius} ({A}raneae: {T}heridiidae), Behav. Ecol. Sociobiol. 17 (1985)
  79--85.

\bibitem{VaKr01}
G.~Vakanas, B.~Krafft, Coordination of behavioral sequences between individuals
  during prey capture in a social spider, \textit{{A}nelosimus eximius}, J.
  Insect Behav. 14 (2001) 777--798.

\bibitem{DeLeCoRoCeOrBo10}
A.~Dejean, C.~Leroy, B.~Corbara, O.~Roux, R.~Céréghino, J.~Orivel, R.~Boulay,
  Arboreal ants use the ``velcro® principle'' to capture very large prey, PLoS
  ONE 5 (2010) 1--7.

\bibitem{PrLiLiMi90}
K.~Pryor, J.~Lindbergh, S.~Lindbergh, R.~Milano, A dolphin-human fishing
  cooperative in {B}razil, Marine Mammal Science 6 (1990) 77--82.

\bibitem{DaCaInLuSi12}
F.~G. Daura-Jorge, M.~Cantor, S.~N. Ingram, D.~Lusseau, P.~C. Simões-Lopes,
  The structure of a bottlenose dolphin society is coupled to a unique foraging
  cooperation with artisanal fishermen, Biol. Lett. 8 (2012) 702--705.

\bibitem{IsRe89}
H.~A. Isack, H.-U. Reyer, Honeyguides and honey gatherers: Interspecific
  communication in a symbiotic relationship, Science 243 (1989) 1343--1346.

\bibitem{SpBeBe16}
C.~N. Spottiswoode, K.~S. Begg, C.~M. Begg, Reciprocal signaling in
  honeyguide-human mutualism, Science 353 (2016) 387.

\bibitem{MiMiLo92}
S.~C. Minta, K.~A. Minta, D.~F. Lott, Hunting associations between badgers
  (\textit{{T}axidea taxus}) and coyotes (\textit{{C}anis latrans}), J. Mammal.
  73 (1992) 814--820.

\bibitem{BsHoAiFr06}
R.~Bshary, A.~Hohner, K.~Ait-el Djoudi, H.~Fricke, Interspecific communicative
  and coordinated hunting between groupers and giant moray eels in the {R}ed
  {S}ea, PLoS Biology 4 (2006) e431.

\bibitem{CrCr95}
S.~Creel, N.~M. Creel, Communal hunting and pack size in african wild dogs,
  \textit{{L}ycaon pictus}, Anim. Behav. 50 (1995) 1325--1339.

\bibitem{Malan98}
G.~Malan, Solitary and social hunting in pale chanting goshawk
  (\textit{{M}elierax canorus}) families: why use both strategies?, J. Raptor
  Res. 32 (1998) 195--201.

\bibitem{BrBa79}
H.~J. Brockmann, C.~J. Barnard, Kleptoparasitism in birds, Anim. Behav. 27
  (1979) 487--514.

\bibitem{VuPeWa04}
J.~A. Vucetich, R.~O. Peterson, T.~A. Waite, Raven scavenging favours group
  foraging in wolves, Anim. Behav. 67 (2004) 1117--1126.

\bibitem{PiMaWi82}
T.~Pitcher, A.~Magurran, I.~Winfield, Fish in larger shoals find food faster,
  Behav. Ecol. Sociobiol. 10 (1982) 149--151.

\bibitem{GaGi01}
B.~G. Galef~Jr., L.~A. Giraldeau, Social influences on foraging in vertebrates:
  causal mechanisms and adaptive functions, Anim. Behav. 61 (2001) 3--15.

\bibitem{Garay09}
J.~Garay, Cooperation in defence against a predator, J. Theor. Biol. 257 (2009)
  45--51.

\bibitem{Pulliam73}
H.~R. Pulliam, On the advantages of flocking, J. Theor. Biol. 38 (1973) 419 --
  422.

\bibitem{Beauchamp08}
G.~Beauchamp, What is the magnitude of the group-size effect on vigilance?,
  Behavioral Ecology 19 (2008) 1361--1368.

\bibitem{PaReLoPeGeJa07}
O.~Pays, P.-C. Renaud, P.~Loisel, M.~Petit, J.-F. Gerard, P.~J. Jarman, Prey
  synchronize their vigilant behaviour with other group members, Proc. R. Soc.
  Lond. B 274 (2007) 1287--1291.

\bibitem{Hamilton71}
W.~Hamilton, Geometry for the selfish herd, J. Theor. Biol. 31 (1971) 295--311.

\bibitem{IoGoCo12}
C.~C. Ioannou, V.~Guttal, I.~D. Couzin, Predatory fish select for coordinated
  collective motion in virtual prey, Science 337 (2012) 1212--1215.

\bibitem{CrQu10}
W.~Cresswell, J.~L. Quinn, Attack frequency, attack success and choice of prey
  group size for two predators with contrasting hunting strategies, Anim.
  Behav. 80 (2010) 643 -- 648.

\bibitem{Giraldeau88}
L.-A. Giraldeau, The stable group and the determinants of foraging group size,
  in: C.~N. Slobodchikoff (Ed.), The Ecology of Social Behavior, Academic
  Press, 1988, pp. 33--53.

\bibitem{Ritz97}
D.~A. Ritz, Costs and benefits as a function of group size: experiments on a
  swarming mysid, \textit{{P}aramesopodopsis rufa} fenton, in: J.~K. Parrish,
  W.~M. Hamner (Eds.), Animal Groups in Three Dimensions, Cambridge University
  Press, 1997, pp. 194--206.

\bibitem{Couzin09}
I.~Couzin, Collective cognition in animal groups, Trends Cogn. Sci. 13 (2009)
  36--43.

\bibitem{CoLi09}
L.~Conradt, C.~List, Group decisions in humans and animals: a survey, Phil.
  Trans. R. Soc. B 364 (2009) 719--742.

\bibitem{BaWa16}
M.~Barbier, J.~R. Watson, The spatial dynamics of predators and the benefits
  and costs of sharing information, PLoS Comput. Biol. 12 (2016) e1005147.

\bibitem{PoSe11}
D.~J. van~der Post, D.~Semmann, Patch depletion, niche structuring and the
  evolution of co-operative foraging, BMC Evolutionary Biology 11 (2011) 335.

\bibitem{PaRu88}
C.~Packer, L.~Ruttan, The evolution of cooperative hunting, Am. Nat. 132 (1988)
  159--198.

\bibitem{LeAuGa04}
C.~Lett, P.~Auger, J.-M. Gaillard, Continuous cycling of grouped vs. solitary
  strategy frequency in a predator-prey model, Theor. Pop. Biol. 65 (2004)
  263--270.

\bibitem{McAuLe06}
R.~Mchich, P.~Auger, C.~Lett, Effects of aggregative and solitary individual
  behaviors on the dynamics of predator-prey game models, Ecol. Model. 197
  (2006) 281.

\bibitem{HoSi98}
J.~Hofbauer, K.~Sigmund, Evolutionary Games and Population Dynamics, Cambridge
  University Press, 1998.

\bibitem{SzFa07}
G.~Szabó, G.~Fáth, Evolutionary games on graphs, Phys. Rep. 446 (2007)
  97--216.

\bibitem{LuCaAr17}
A.~F. Lütz, A.~Cazaubiel, J.~J. Arenzon, Cyclic competition and percolation in
  grouping predator-prey populations, Games 8 (2017) 10.

\bibitem{AnSc06}
T.~Antal, I.~Scheuring, {Fixation of strategies for an evolutionary game in
  finite populations}, Bull. Math. Biol. 68 (2006) 1923--1944.

\bibitem{ReMoFr07}
T.~Reichenbach, M.~Mobilia, E.~Frey, Mobility promotes and jeopardizes
  biodiversity in rock-paper-scissors games, Nature 448 (2007) 1046--1049.

\bibitem{LuRiAr13}
A.~F. Lütz, S.~Risau-Gusman, J.~J. Arenzon, Intransitivity and coexistence in
  four species cyclic games, J. Theor. Biol. 317 (2013) 286--292.

\bibitem{SzMoJiSzRuPe14}
A.~Szolnoki, M.~Mobilia, L.-L. Jiang, B.~Szczesny, A.~M. Rucklidge, M.~Perc,
  Cyclic dominance in evolutionary games: a review, J. R. Soc. Interface 11
  (2014) 20140735.

\bibitem{VeGaAv10}
M.~van Veelen, J.~García, L.~Avilés, It takes grouping and cooperation to get
  sociality, J. Theor. Biol. 264 (2010) 1240--1253.

\bibitem{PoLe14}
J.~R. Potts, M.~A. Lewis, How do animal territories form and change? {L}essons
  from 20 years of mechanistic modelling, Proc. R. Soc. Lond. B 281 (2014)
  20140231.

\bibitem{Read16}
J.~E. Herbert-Read, P.~Romanczuk, S.~Krause, D.~Strömbom, P.~Couillaud,
  P.~Domenici, R.~H. J.~M. Kurvers, S.~Marras, J.~F. Steffensen, A.~D.~M.
  Wilson, J.~Krause, Proto-cooperation: Group hunting sailfish improve hunting
  success by alternating attacks on grouping prey, Proc. R. Soc. B 283 (2016)
  20161671.

\bibitem{KrRe96}
P.~L. Krapivsky, S.~Redner, Kinetics of a diffusive capture process: Lamb
  besieged by a pride of lions, J. Phys. A: Math. Gen. 29 (1996) 5347--5357.

\bibitem{Nahin07}
P.~J. Nahin, Chases and Escapes: The Mathematics of Pursuit and Evasion,
  Princeton University Press, 2007.

\bibitem{CoPlBe08}
E.~A. Codling, M.~J. Plank, S.~Benhamou, Random walk models in biology, J. R.
  Soc. Interface 5 (2008) 813--834.

\bibitem{GiBa10}
L.~Giuggioli, F.~Bartumeus, Animal movement, search strategies and behavioural
  ecology: a cross-disciplinary way forward, J. Anim. Ecol. 79 (2010) 906--909.

\bibitem{RaBaLuRiSoVi11}
E.~P. Raposo, F.~Bartumeus, M.~G.~E. da~Luz, P.~J. Ribeiro-Neto, T.~A. Souza,
  G.~M. Viswanathan, How landscape heterogeneity frames optimal diffusivity in
  searching processes, PLoS Comp. Biol. 7 (2011) e1002233.

\bibitem{Angelani12}
L.~Angelani, Collective predation and escape strategies, Phys. Rev. Lett. 109
  (2012) 118104.

\bibitem{ViZa12}
T.~Vicsek, A.~Zafeiris, Collective motion, Phys. Rep. 517 (2012) 71--140.

\bibitem{VaSiAr07}
M.~H. Vainstein, A.~T.~C. Silva, J.~J. Arenzon, {Does mobility decrease
  cooperation?}, J. Theor. Biol. 244 (2007) 722--728.

\bibitem{ScKi13}
S.~J. Schreiber, T.~P. Killingback, Spatial heterogeneity promotes coexistence
  of rock-paper-scissors metacommunities, Theor. Pop. Biol. 86 (2013) 1--11.

\bibitem{BhVi14}
K.~Bhattacharya, T.~Vicsek, Collective foraging in heterogeneous landscapes, J.
  R. Soc. Interface 11 (2014) 20140674.

\bibitem{BhVi15}
K.~Bhattacharya, T.~Vicsek, To join or not to join: collective foraging
  strategies, J. Phys.: Conf. Series 638 (2015) 012015.

\bibitem{SaTo94}
J.~E. Satulovsky, T.~Tomé, Stochastic lattice gas model for a predator-prey
  system, Phys. Rev. E 49 (1994) 5073--5079.

\bibitem{KuAb01}
M.~Kuperman, G.~Abramson, Small world effect in an epidemiological model, Phys.
  Rev. Lett. 86 (2001) 2909--2912.

\bibitem{RuAr14}
C.~Rulquin, J.~J. Arenzon, Globally synchronized oscillations in complex cyclic
  games, Phys. Rev. E 89 (2014) 032133.

\end{thebibliography}

\end{document}